\documentclass[
 reprint,
superscriptaddress,
 amsmath,amssymb,
 aps,
]{revtex4-1}
\usepackage{color}
\usepackage{graphicx}
\usepackage{caption}
\usepackage{subcaption}
\usepackage{dcolumn}
\usepackage{bm}
\usepackage{braket}
\usepackage{float}
\usepackage{multirow}

\begin{document}


\title{Direct measurement of Bacon-Shor code stabilizers}

\author{Muyuan Li}
\affiliation{School of Computational Science and Engineering,
 Georgia Institute of Technology, Atlanta, Georgia 30332, USA}
\author{Daniel Miller}
\affiliation{Theoretical Physics $\MakeUppercase{\romannumeral3}$, Heinrich Heine University Duesseldorf, D-40225 Duesseldorf, Germany}
\author{Kenneth R. Brown}
\affiliation{School of Computational Science and Engineering,
 Georgia Institute of Technology, Atlanta, Georgia 30332, USA}
\affiliation{Departments of Electrical and Computer Engineering, Chemistry, and Physics, Duke University, Durham, NC 27708, USA}

\email{ken.brown@duke.edu}

\begin{abstract}
A Bacon-Shor code is a subsystem quantum error-correcting code on an $L \times L$ lattice where the $2(L-1)$ weight-$2L$ stabilizers are usually inferred from the measurements of $(L-1)^2$ weight-2 gauge operators. Here we show that the stabilizers can be measured directly and fault tolerantly with bare ancillary qubits by constructing circuits that follow the pattern of gauge operators.  We then examine the implications of this method for small quantum error-correcting codes by comparing distance 3 versions of the rotated surface code and the Bacon-Shor code with the standard depolarizing model and in the context of a trapped ion quantum computer. We find that for a simple circuit of prepare, error correct and measure the Bacon-Shor code outperforms the surface code by requiring fewer qubits, taking less time, and having a lower error rate.    

\end{abstract}

\pacs{Valid PACS appear here}
\maketitle

Quantum information experiments are approaching the number of qubits and operational fidelity necessary for quantum error correction to improve performance \cite{BermudezITQCwithSteane2017,trout2017simulating,OBrienSurface17DensityMatrix2017}. Classical error correction on quantum devices have already shown the ability to suppress introduced errors and increase memory times \cite{kelly2015state, taminiau2014universal, NMR3, chiaverini2004realization, SchindlerRepCpde2011}. Two promising quantum error-correcting codes for data qubits arranged on an $L\times L$ lattice are the surface code \cite{SCorig, Tilted13SC, TomitaLowDSC2014} and the Bacon-Shor code \cite{BaconBaconShor2006, ShorBaconShor1995, aliferis2007subsystem}.  Numerical simulation of the surface code shows a high memory threshold of 1\% error per operation for increasing $L$ and for distance 3 codes a pseudothreshold of 0.3\% error per operation for a depolarizing error model \cite{TomitaLowDSC2014}.  The Bacon-Shor code is a subsystem code and has no threshold as $L$ grows \cite{NappOptimalBaconShor} but promising performance for small distance codes with a pseudothreshold of 0.2\% for a depolarizing error model \cite{aliferis2007subsystem} and a fault-tolerant protocol for implementing universal gates without distillation \cite{YoderBS2017}. The rotated surface code has $L^2-1$ check operators of weight 4 in the bulk and weight 2 on the boundary \cite{Tilted13SC}.  The advantage of the Bacon-Shor code comes from using weight-2 gauge operators to determine the weight 2$L$ check operators and the lack of threshold is a result of having only $2(L-1)$ checks \cite{BaconBaconShor2006, aliferis2007subsystem, NappOptimalBaconShor}. 

For the [[9,1,3]] surface code, Tomita and Svore \cite{TomitaLowDSC2014} pointed out that in the circuit model, the order in which the weight 4 check operators are measured prevents unwanted error propagation and allows for parallelization of operations. For Bacon-Shor, the idea has always been to measure gauge operators because direct measurements of the stabilizers would require long-range interaction between qubits and higher-weight stabilizers usually require more complicated ancillary qubit preparation \cite{ShorCatState, SteaneSteaneEC1997, KnillKnillEC2005, ChaoFlagQubits12017}. Inspired by Tomita and Svore and the fact that both the surface code and the Bacon-Shor code can be thought as gauge choices on the compass model, we found that Bacon-Shor codes can be measured with bare ancillary qubits.  

\begin{figure*}
    \centering
    \begin{subfigure}[b]{0.2\textwidth}
        \centering
        \includegraphics[width=\linewidth]{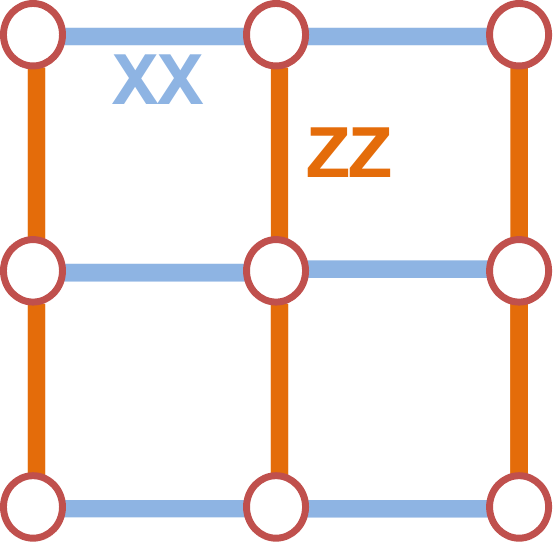}
        \caption{Compass Model}
    \label{fig:compass_compass}
    \end{subfigure}%
    ~~~~~~~~~~~~~
    \begin{subfigure}[b]{0.2\textwidth}
        \centering
        \includegraphics[width=\linewidth]{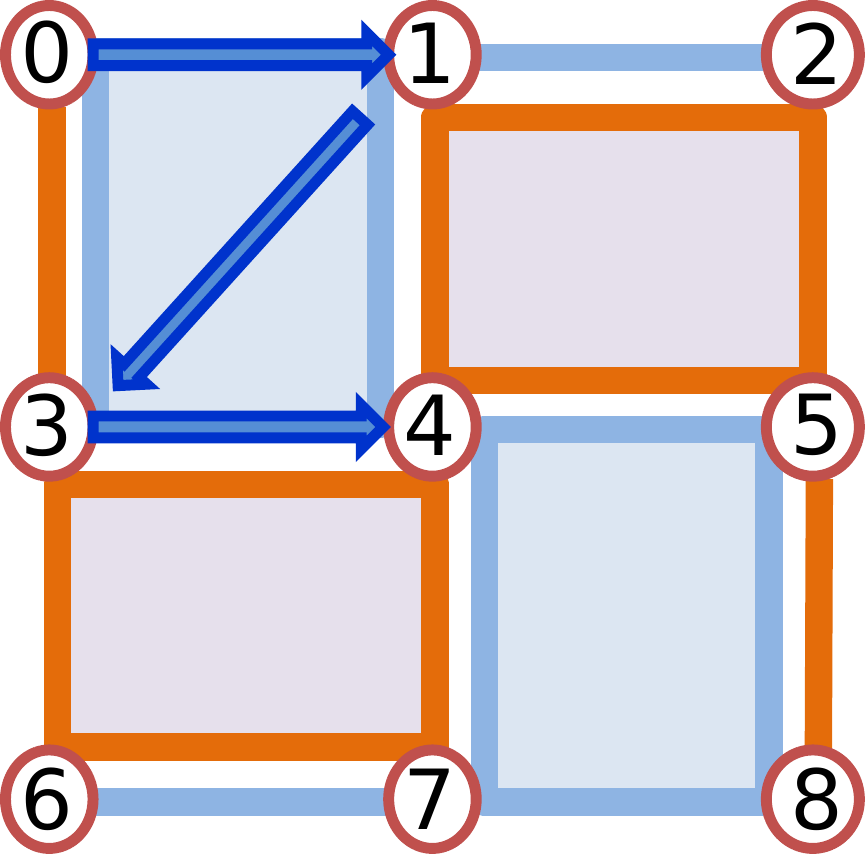}
        \caption{Surface}
    \label{fig:compass_surface}
    \end{subfigure}
    ~~~~~~~~~~~~~
    \begin{subfigure}[b]{0.2\textwidth}
        \centering
        \includegraphics[width=\linewidth]{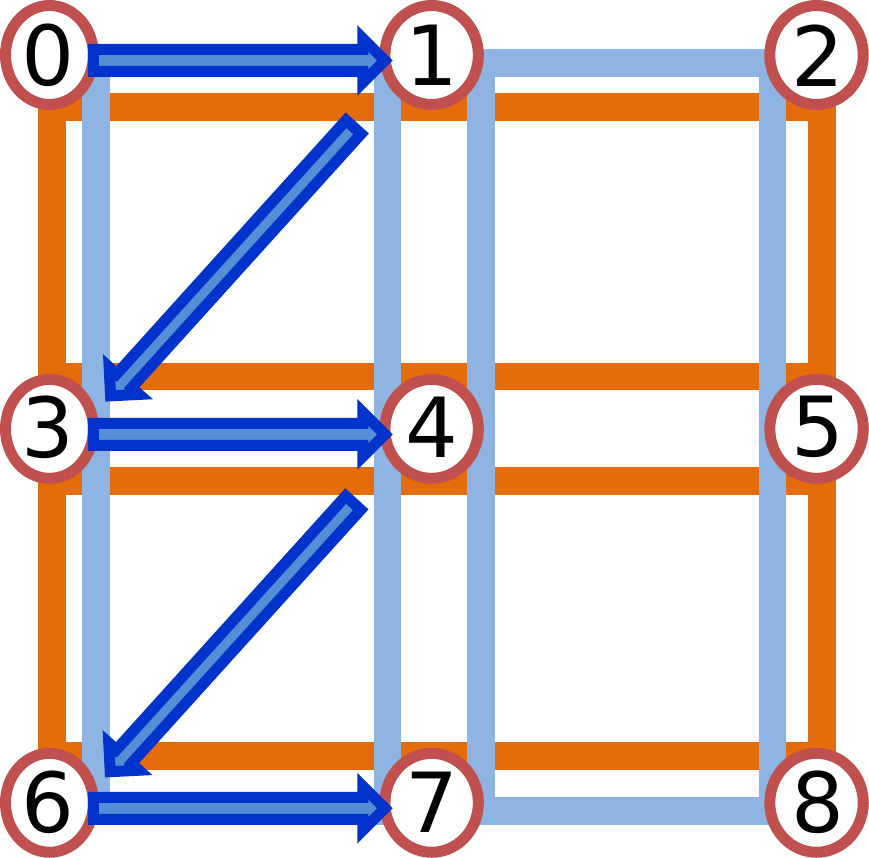}
        \caption{Bacon-Shor}
    \label{fig:compass_BS}
    \end{subfigure}
    \captionsetup{justification=raggedright}
    \caption{The compass model with $ZZ$ bonds along the vertical axis and $XX$ bonds along the horizontal axis. Choices of gauge on a $3 \times 3$ lattice lead to two well-known stabilizer codes: [[9,1,3]] surface code and [[9,1,3]] Bacon-Shor code . The underlying bonds of the compass model are a guide for how to fault tolerantly measure surface code and Bacon-Shor code stabilizers with bare ancillary qubits. Measuring stabilizers in order of gauge operators can help suppress hook errors on two-qubit gates in the stabilizer measurement circuit. The blue arrows show the circuit order for measuring an $X$-type stabilizer for both codes.}
\label{fig:compass}
\end{figure*}
In condensed matter physics the compass model is used to describe a family of lattice models involving interacting quantum degrees of freedom \cite{nussinov2015compass}. The relationship between compass model and topological quantum computing is also well-studied, with example such as the Kitaev's honeycomb model \cite{KITAEV2006}. FIG. \ref{fig:compass_compass} shows an example of a 9-qubit compass model on square lattice with Pauli type interactions between neighboring qubits.

Subsystem error-correcting codes arise naturally from the compass model, where the interactions between neighboring qubits can be viewed as weight-$2$ gauge operators defined by the subsystem code \cite{BaconBaconShor2006}. The compass model also has a $90^{\circ}$ rotation symmetry, so that $X$ and $Z$ errors are treated symmetrically. Examining FIG. \ref{fig:compass} we see that the rotated surface code and the Bacon-Shor code can be considered as different choices of gauge fixings on this compass model: for the surface code, each weight-2 stabilizer is exactly a gauge operator of the corresponding type, and each weight-4 stabilizer is equivalent to fixing the parity of the product of two gauge operators on the same face; for the Bacon-Shor code, each weight-6 stabilizer is equivalent to fixing the parity of the product of three gauge operators in the same double rows or columns of qubits.

The surface code is considered a promising candidate for fault-tolerant quantum computing \cite{RaussendorfClusterState22007, FowlerSurfaceCodeThresh2009, horsman2012surface}. It is also a popular choice for implementing error correction on near-term small quantum devices \cite{fowler2012surface, versluis2017scalable, trout2017simulating} due to its ability to restrict all stabilizer measurements as local operations and to perform fault-tolerant syndrome extraction with a bare ancillary qubit per check operator. The ability to use bare ancillary qubits relies on a proper choice of circuit for implementing syndrome measurement of the [[9,1,3]] surface code, as illustrated in Fig.\ref{fig:compass_surface}. This choice has been previously described as directing the hook errors away from the direction of the logical error \cite{TomitaLowDSC2014}.

Considering the surface code from the perspective of compass model, the order of performing two-qubit gates in the stabilizer measurement circuit follows from the condition that the stabilizers are built from the gauges, as the weight-2 edge check operators correspond directly to interactions on the compass model, and the bulk check operators correspond to product of two interactions. By measuring each stabilizer in the order of individual interactions on the compass model, the measurement prevents single errors on ancillary qubits from propagating to two errors on the data (hook errors) that contribute to a logical error. This procedure is critical for the distance-3 surface code to be fault tolerant to single qubit errors. For higher distance codes, it prevents weight-2 errors along the direction of the logical operator from occurring with the probability of a weight-1 error. 

The two-dimensional Bacon-Shor code on a $L \times L$ lattice is consisted of $L-1$ number of double columns of $X$ stabilizers and $L-1$ number of double rows of $Z$ stabilizers. As a subsystem code, the Bacon-Shor code can be defined by gauge algebra. The gauge group of the code is generated by two-qubit $XX$ operators acting on neighboring qubits in the same row and two-qubit $ZZ$ operators acting on neighboring qubits in the same column. The logical $X$ operators are $X^{\otimes L}$ acting on all qubits in the same column, and logical $Z$ operators are $Z^{\otimes L}$ acting on all qubits in the same row.

Suppose we consider only $X$ errors on the lattice, then an even number of $X$ errors on the same row can be seen as a product of $XX$ gauge operators and thus does not affect the logical state of the code. Thus for each row we only need to consider the parity of the number of $X$ errors. A similar statement can be made for $Z$ errors on the lattice.

To obtain the parity values of the $(L-1)^2$ stabilizers one can measure the $L \times (L-1)$ $XX$ gauge operators and the $L \times (L-1)$ $ZZ$ gauge operators, and use this information to find the eigenvalues of the $X$ and $Z$ stabilizers. The advantage of this method is that the operators are fully local, where ancillary qubits can be placed in between the two data qubits being measured. The weight-$2L$ parity checks can also be measured directly by preparing a $2L$-qubit GHZ state. The problem with both of these methods is that they require a large amount of ancillary qubits, and this makes the Bacon-Shor code inferior to the surface code when implemented on small quantum devices. However, by considering gauges and using insights from Surface-17 on the compass model, a proper circuit for measuring the stabilizers can be chosen so that the weight-$2L$ stabilizers can be directly measured fault tolerantly using a single bare ancillary qubit.

The challenge of fault-tolerant $k$-weight stabilizer measurement is that errors on the ancillary qubit can generate hook errors on the data of weight $k/2+1$.  By measuring the stabilizers following the structure of the gauge operators, the hook errors are simply products of gauges and a single qubit error. Topologically one can consider the sequential product of gauge errors as a string with both ends attached to the same boundary and is therefore a trivial operator. This is in contrast to a logical operator where the string connects opposite boundaries \cite{SCorig}.

For the $Z$ stabilizers, the circuit consists of preparing the ancillary qubits in $\ket{0}$ and then performing $2L$ controlled-not gates with each data ion as the control and the ancillary ion as the target, followed by measurement in the $Z$ basis.  We order the controlled-not gates such that the target qubits come in pairs that follows the $ZZ$ gauge operators. A similar order holds for the $X$ stabilizers, where now the ancilla is prepared and measured in the $X$ basis and the controlled-not targets the data qubits. This circuit has already been implemented experimentally in trapped ions and superconducting qubits for the $L=2$ Bacon-Shor quantum error detection code \cite{Linke422Ions2016,TakitaPRL2017}

 As an example for the distance-3 code, the order of controlled-not gates for the two $Z$ stabilizers is 
\begin{enumerate}
    \item $Z_0\rightarrow Z_3\rightarrow Z_1\rightarrow Z_4\rightarrow Z_2\rightarrow Z_5, $
    \item $Z_3\rightarrow Z_6\rightarrow Z_4\rightarrow Z_7\rightarrow Z_5\rightarrow Z_8.$
\end{enumerate}
By using a similar order for the $X$ stabilizers in terms of $XX$ gauges, the order for the distance-3 code is 
\begin{enumerate}
    \item $X_0\rightarrow X_1\rightarrow X_3\rightarrow X_4\rightarrow X_6\rightarrow X_7,$
    \item $X_1\rightarrow X_2\rightarrow X_4\rightarrow X_5\rightarrow X_7\rightarrow X_8,$
\end{enumerate}
fault-tolerant syndrome measurement can be achieved for all stabilizers.

With this specific measurement order we only need one ancillary qubit per stabilizer, and can perform fault-tolerant syndrome measurement of the [[9,1,3]] qubit Bacon-Shor code. Instead of reusing ancillary qubits, we choose to use one per syndrome measurement yielding a total of $13$ qubits. We refer to this choice as Bacon-Shor-13 following the notation of Tomita and Svore where the [[9,1,3]] surface code with 8 ancilla is referred to as Surface-17 \cite{TomitaLowDSC2014}.

Syndrome measurement is not the whole process of quantum error correction. Another important part is the preparation of logical states. For the surface code, to encode logical $\ket{0}$ we prepare all the data qubits in the physical $\ket{0}$ state, measure $X$ type stabilizer 2-3 times and perform correction based on the syndrome.  For Bacon-Shor-13, to encode logical $\ket{0}$ we simply prepare three 3-qubit GHZ states in the $X$ basis down the columns without verification, $\ket{0}_L=\bigotimes^2_{i=0} (\frac{1}{\sqrt{2}}\ket{+++}+\ket{---})_{0+i,3+i,6+i}$. The preparation is fault tolerant and deterministic. To compare the performance of Bacon-Shor-13 and Surface-17 we focus on simulating a circuit with 3 elements: logical state encoding, quantum error correction, and measurement of the individual data qubits. We refer to this circuit as the simple circuit. Using the measurement results, we determine the outcome of the logical circuit and the probability that the circuit fails.

To perform error correction using Surface-17 and Bacon-Shor-13, we designed two-step lookup table decoders for both codes. The details of the Surface-17 decoder can be found at \cite{trout2017simulating}. For any two-step decoder, in the first step if the syndrome shows no errors then no correction is performed; if the syndrome shows errors, then a second syndrome is measured and correction is applied based on the second syndrome. All simulations are performed with CHP \cite{aaronson2004improved} using importance sampling. The importance sampling method is described in \cite{LiBareAnc2017, trout2017simulating}.

\begin{figure}[ht]
\centering
    \begin{subfigure}[b]{0.24\textwidth}
        \centering
        \includegraphics[width=\linewidth]{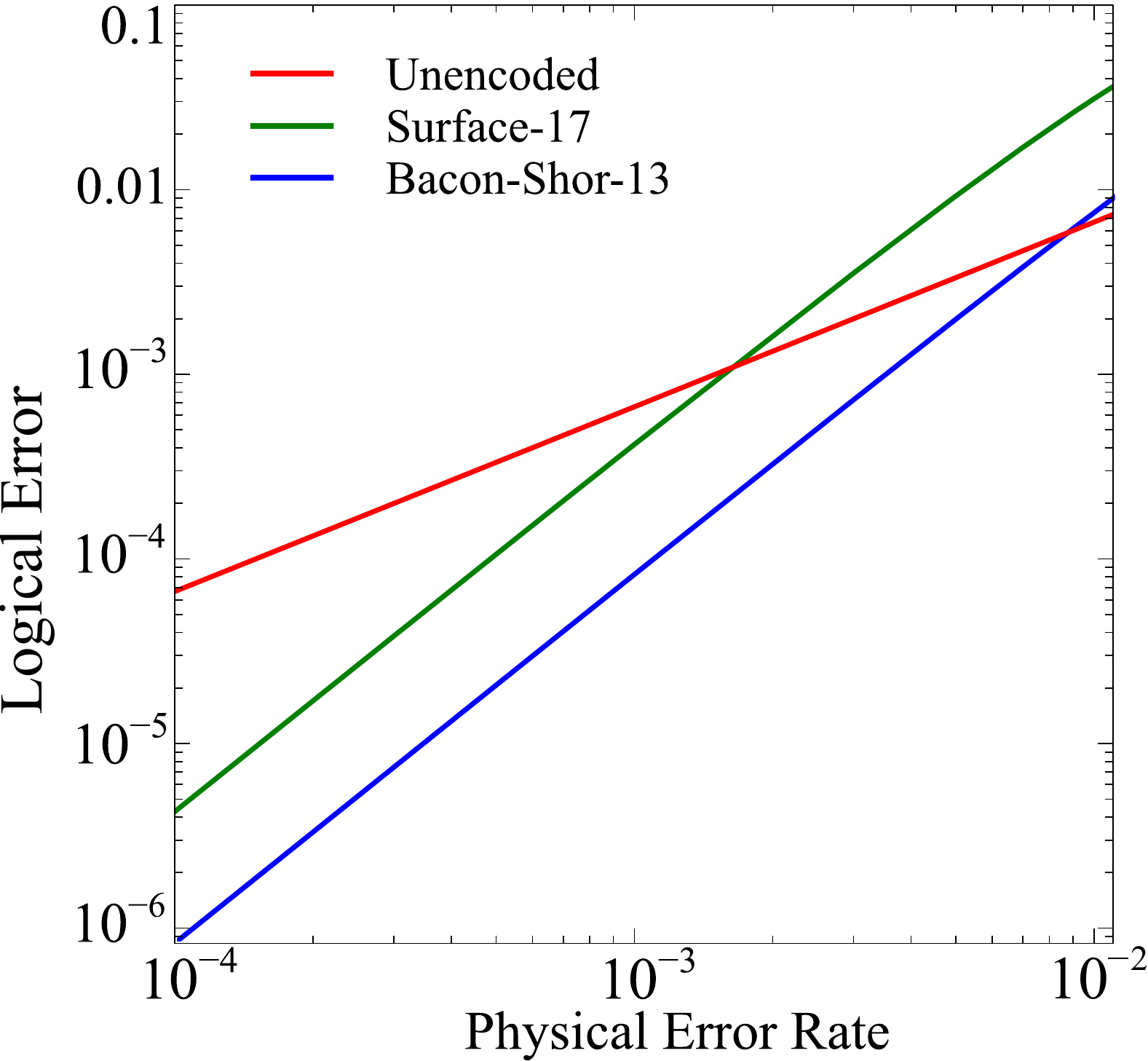}
        \caption{One round of error correction}
    \label{fig:simple_one}
    \end{subfigure}%
    \hfill%
    \begin{subfigure}[b]{0.23\textwidth}
        \centering
        \includegraphics[width=\linewidth]{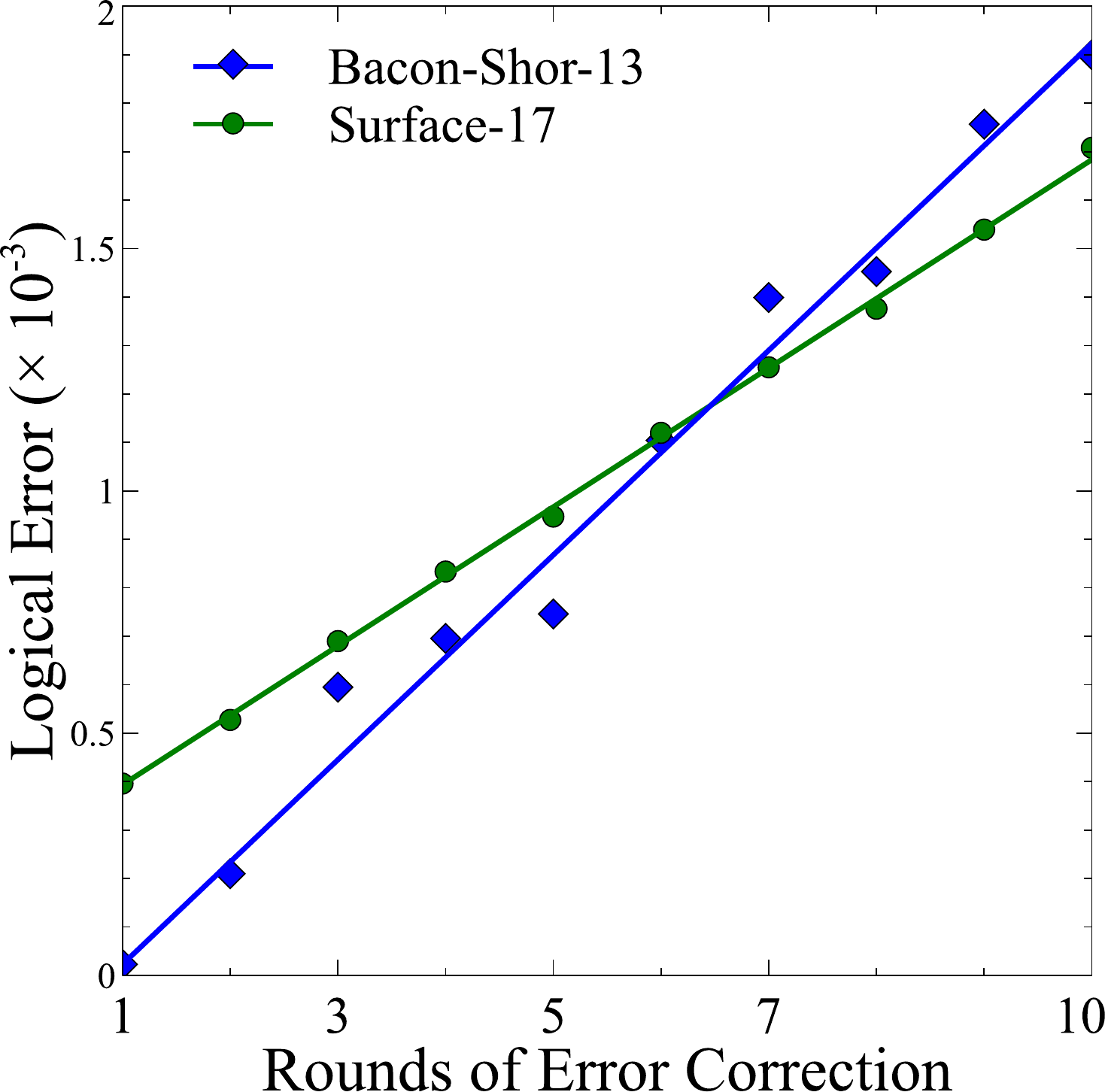}
        \caption{Multiple rounds of error correction}
    \label{fig:Simple_mult}
    \end{subfigure}
\captionsetup{justification=raggedright}
\caption{Comparison of Bacon-Shor-13 and Surface-17 in a simple circuit simulation. (a) With one round of error correction Bacon-Shor-13 shows a pseudothreshold of $0.9$\% and Surface-17 shows a pseudothreshold of $0.15$\%. Such difference is mainly due to the difference in logical state preparation. (b) At error rate of $10^{-3}$ Surface-17 starts to outperform Bacon-Shor-13 in a simple circuit with more than $7$ rounds of error correction.}
\label{fig:standard}
\end{figure}

As a standard error model used in simulations of quantum error-correcting codes, the depolarizing model is used to benchmark the performance of Bacon-Shor-13 and Surface-17. For each gate in the circuit, an element from the one-qubit (two-qubit) Pauli group is sampled and applied after the gate to serve as an error. For simulations of the simple circuit we assume all one-qubit, two-qubit gates and measurements have the same error rate.

As Surface-17 requires a large overhead of operations for logical state preparation when compared to Bacon-Shor-13, the fault-tolerant performance of the two codes will be largely dependent on the performance of state preparation. When only one round of error correction is performed, although Surface-17 can correct more errors than Bacon-Shor-13, the overhead of state preparation makes its overall performance inferior to Bacon-Shor-13, as illustrated in FIG. \ref{fig:simple_one}. When multiple rounds of error correction are executed, Surface-17's advantage in error correction starts to dominate. As shown in FIG. \ref{fig:Simple_mult} Surface-17 starts to outperform Bacon-Shor-13 in a simple circuit with more than $7$ rounds of error correction.

To simulate the performance of error-correcting codes on an ion trap quantum computer we need to map the codes onto a linear ion chain and compile controlled-not gates from M{\o}lmer-S{\o}rensen gates \cite{MolmerPRL1999,MaslovCircuitCompIT2017}. Details of operations with ion trap quantum computer can be found in \cite{trout2017simulating}. Using a simulated annealing algorithm, we searched for ion chain arrangements that minimized total time for quantum error correction or average two-qubit gate time (see Appendix). In our ion trap error model, gate error scales with two-qubit gate time and minimizing average two-qubit gate time minimizes the error.  Times required to execute the simple circuit when the ions are arranged to minimize the average two-qubit gate time are shown in TABLE \ref{table:times}. Bacon-Shor-13 holds clear advantage in terms of circuit execution time.  These reported times are not fundamental and can be improved by changes in gate and measurement schemes \cite{LeungFMgates2017,BermudezITQCwithSteane2017} but for all protocols, Bacon-Shor-13 will maintain the time advantage of deterministic state preparation over Surface-17. 

\begin{table}
\begin{tabular}{ c|c|c|c|c }
 \hline
 Code & Prep & QEC & Measure & Total \\ \hline \hline
Surface-17 & 2400-3600 & 4900-9800 & 100 & 7400-13500\\ \hline
Bacon-Shor-13 & 1670 & 4310-8620 & 100 & 6080-10390\\ \hline
\end{tabular}
\captionsetup{justification=raggedright}
\caption{Time required (in $\mu$s) to execute a simple circuit of prepare, error correct, and measure using an ion trap model (see \cite{trout2017simulating} and Appendix). The required times can vary as the rounds of stabilizer measurements depends on the results of syndrome measurements.}
\label{table:times}
\end{table}

\begin{figure*}
    \centering
    \begin{subfigure}[b]{0.34\textwidth}
        \centering
        \includegraphics[width=\linewidth]{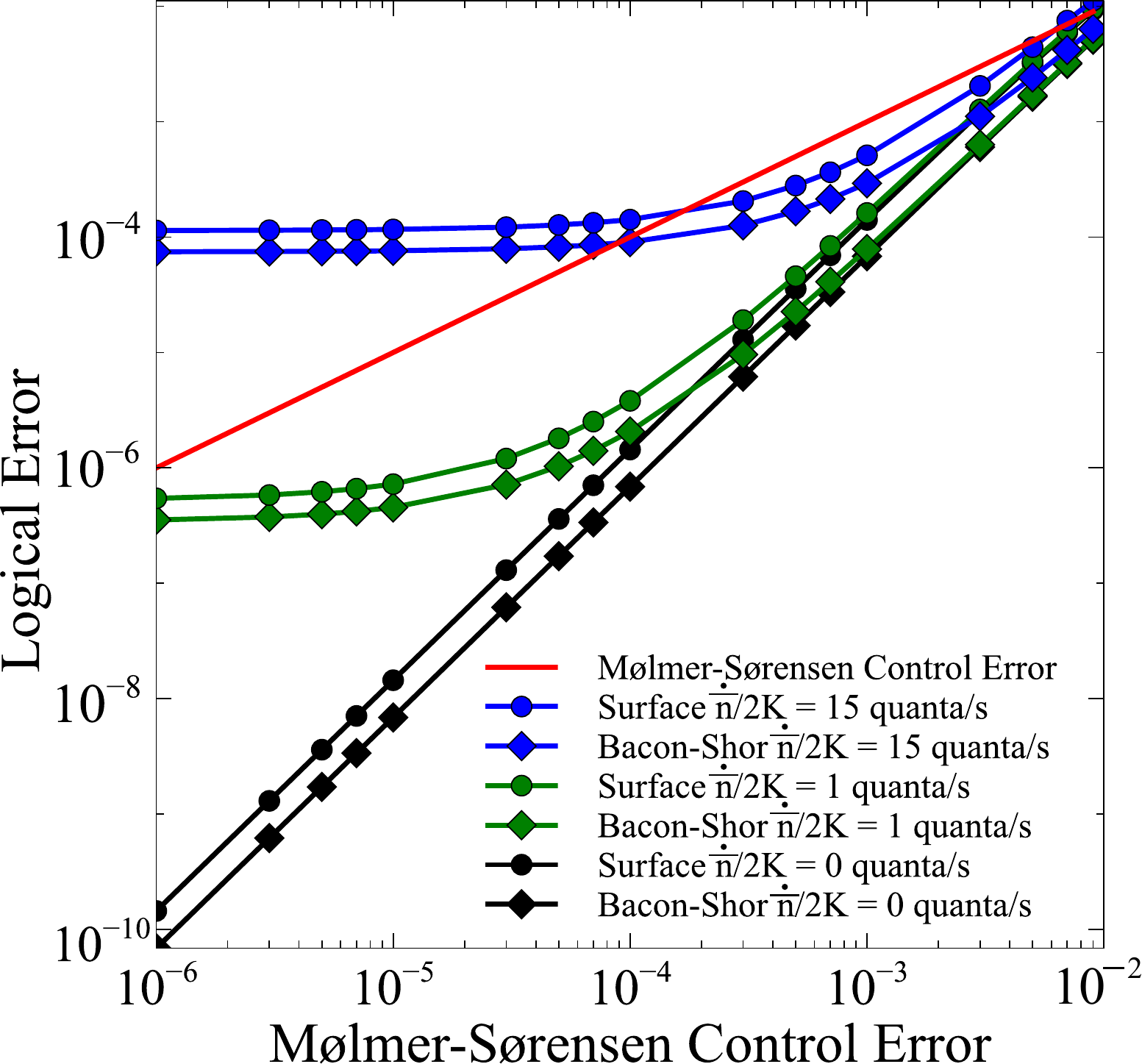}
        \caption{Ion heating error}
    \end{subfigure}%
    ~~~~~~~~~~~~~~~~~~~~
    \begin{subfigure}[b]{0.34\textwidth}
        \centering
        \includegraphics[width=\linewidth]{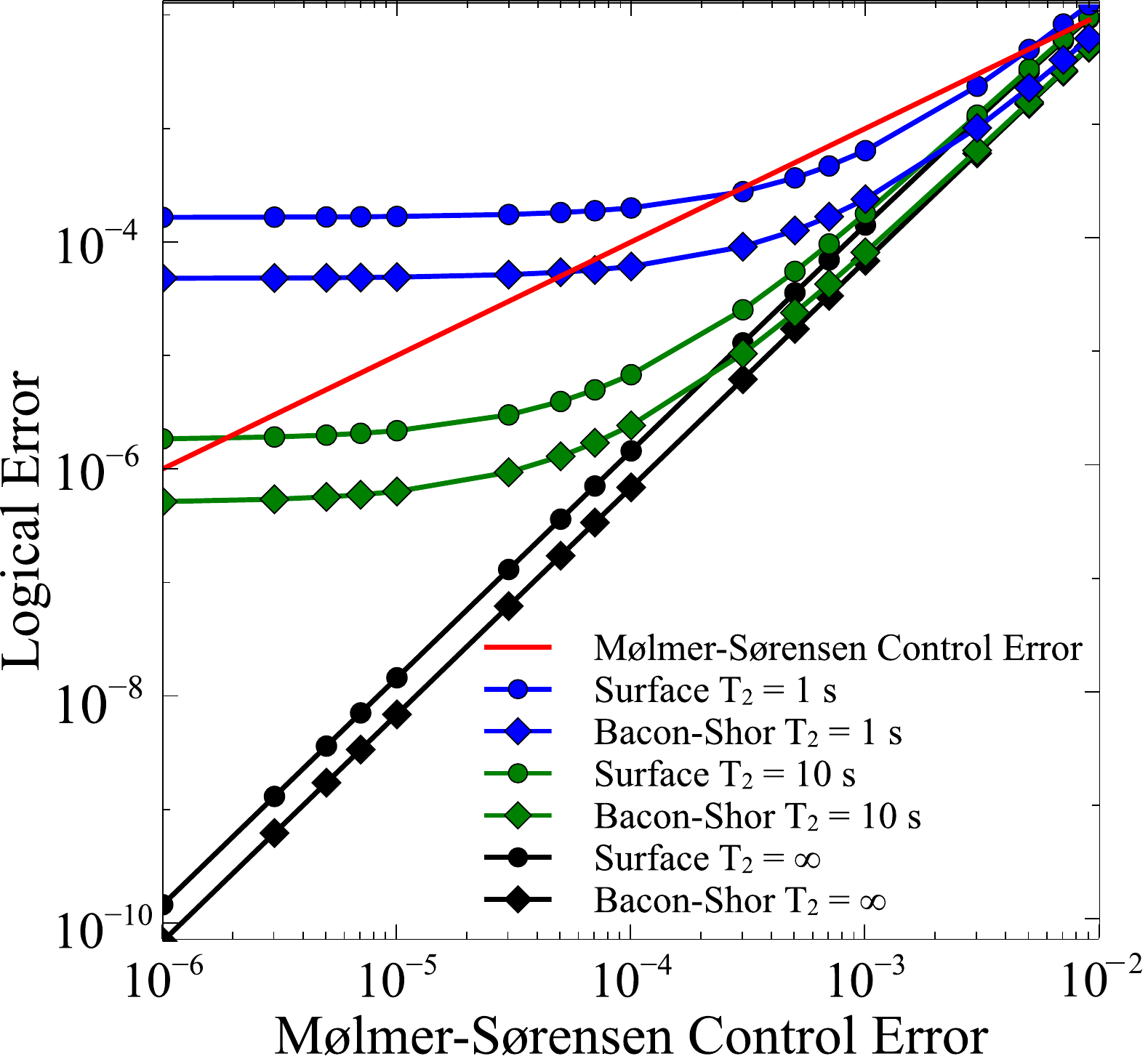}
        \caption{Spin dephasing error}
    \end{subfigure}
    \captionsetup{justification=raggedright}
    \caption{Comparison of Bacon-Shor-13 and Surface-17 with a simple circuit under the influence of different ion trap error sources. In each plot, in addition to over rotations of M{\o}lmer-S{\o}rensen gates only the labeled error source was introduced in the simulation. Bacon-Shor-13 yields a lower logical error rate than Surface-17 for any error sources and any error strength.}
\label{fig:iontrap}
\end{figure*}

We use the error model as described in detail in \cite{trout2017simulating} to simulate error sources in an ion trap quantum computer. A summary of the error model is presented in the Appendix. Note that the most significant error sources are coherent over-rotations of the M{\o}lmer-S{\o}rensen gates, motional mode heating of ions, and dephasing errors generated by AC stark effect. The error probabilities of motional mode heating and dephasing errors depend on the operation time of the corresponding gate.

In FIG. \ref{fig:iontrap} we present the results of comparing Bacon-Shor-13 and Surface-17 in a simple circuit with one round of error correction under the influence of different ion trap error sources. Here we assume the error rate for single-qubit gates and measurements is ten times smaller than for two-qubit gates in order to match realistic error rates in experiments. All results are computed for an optimization of ion chain arrangement that minimizes average 2-qubit gate times for both codes independenlty (see Appendix).

From FIG. \ref{fig:iontrap} we notice that for all error sources and all error strengths Bacon-Shor-13 outperforms Surface-17. For both codes, the simulated simple circuit would outperform a single two-qubit gate assuming only M{\o}lmer-S{\o}rensen control errors with error rate below $10^{-2}$. We also see that for a heating rate of 30 quanta/s, both codes can barely outperform a single two-qubit gate at error rate of $3\times 10^{-3}$. The heating rate error follows the model of \cite{BallanceControlHFQubits2016}. This indicates that a two-qubit gate error rate of $10^{-3}$ should clearly demonstrate that an encoded circuit outperforms the unencoded circuit.

In conclusion, we have shown that Bacon-Shor-13 outperforms Surface-17 in all measures for this simple circuit: time, logical error rate, and number of qubits. The key advantage of Bacon-Shor-13 over Surface-17 comes from its greatly simplified state preparation. In addition, the lower qubit count makes Bacon-Shor-13 a more immediate target for near-term quantum error correction in systems where non-nearest neighbor gates are possible, such as trapped ions. However as Surface-17 holds advantage over Bacon-Shor-13 in terms of error correction, multiple rounds of error correction will begin to favor Surface-17.

We note that the compass model code on an $L\times L$ lattice allows for a family of codes defined by how the gauge operators are fixed and the Bacon-Shor and rotated surface code are two extremes of this family. We are currently studying this family of codes to determine what choice of gauges would yield a threshold, as the codes transition from Bacon-Shor to surface code \cite{michael2018notes}.

We thank Dave Bacon, Andrew Cross, Cody Jones, and Colin Trout for useful discussions. This work was supported
by the Office of the Director of National Intelligence - Intelligence Advanced Research Projects Activity through ARO contract W911NF-10-1-0231, the ARO MURI on Modular
Quantum Systems W911NF-16-1-0349, the National Science Foundation grant PHY-1415461, and the Alexander von Humboldt Foundation. 
\bibliographystyle{apsrev}
\bibliography{References}

\section*{Appendix}
\begin{table*}
    \centering
    \begin{tabular}{c|c|c}
    \hline
    Code & Opt. & Ion Arrangement \\ \hline \hline
    \multirow{3}{*}{Surface-17} & SA & 0 \, 2 \, 6 \, 8 \, 1 \, 4 \, 3 \, 7 \, 5 \, \textbf{11 \, 12 \, 10 \, 15 \, 13 \, 14 \, 9 \, 16}\\
     & MA & 2 \, \textbf{9} \, 1 \, \textbf{12} \, 5 \, \textbf{15} \, 8 \, \textbf{14} \, 4 \, \textbf{11} \, 0 \, \textbf{10} \, 3 \, \textbf{13} \, 7 \, \textbf{16} \, 6\\
     & MT & \textbf{10} \, \textbf{15} \, \textbf{9} \, 5 \, 0 \, 1 \, \textbf{11} \, \textbf{12} \, \textbf{14} \, 7 \, 4 \, 3 \, 8 \, 2 \, 6 \, \textbf{13} \, \textbf{16}\\ \hline
     \multirow{3}{*}{Bacon-Shor-13} & SA & 0 \, 2 \, 6 \, 8 \, 1 \, 3 \, 7 \, 5 \, 4 \, \textbf{11 \, 10 \, 12 \, 9} \\
     & MA & 8 \, 2 \, \textbf{12} \, 1 \, 5 \, \textbf{9} \, 4 \, \textbf{10} \, 7 \, 3 \, \textbf{11} \, 0 \, 6 \\
     & MT & 2 \, 1 \, 5 \, 4 \, \textbf{9 \, 12 \, 10 \, 11} \, 7 \, 3 \, 0 \, 8 \, 6\\ \hline
    \end{tabular}
    \captionsetup{justification=raggedright}
    \caption{Ion arrangements optimized for an array of parameters (see text). Numbers in bold face represent ancillary qubits. }
    \label{table:arrangement}
\end{table*}
\subsection{Kraus channels for error models}
In our simulations, we use both a standard depolarizing error model and an ion trap inspired Pauli error model~\cite{trout2017simulating}. 
\begin{enumerate}
    \item \textbf{Standard Error Model}\\
    The one- and two-qubit Kraus channels are of the form
    \begin{equation}
    \begin{aligned}
      E_1 &= \{\sqrt{1-p}I, \sqrt{\frac{p}{3}}X, \sqrt{\frac{p}{3}}Y, \sqrt{\frac{p}{3}}Z\}, \\
      E_2 &= \{\sqrt{1-p}II, \sqrt{\frac{p}{15}}IX, \ldots, \sqrt{\frac{p}{15}}ZZ\},
    \end{aligned}
    \end{equation}
    where $p$ is the error rate of the error channel. For each gate in the circuit, an element from the one-qubit (two-qubit) Pauli group is sampled and applied after the gate (before for measurements) to serve as an error.

    \item \textbf{Ion Trap Error Model}\\
    A \textit{CNOT} gate is constructed from a M{\o}lmer-S{\o}rensen entangling gate \textit{XX} and single qubit rotations \textit{RX}, \textit{RY} \cite{MaslovCircuitCompIT2017}. The Kraus channels for these gates are
    \begin{equation}
    \begin{aligned}
      E_{XX} &= \{\sqrt{1-p_{XX}}II, \sqrt{p_{XX}}XX\}, \\
      E_{RX} &= \{\sqrt{1-p_{RX}}I, \sqrt{p_{RX}}X \}, \\
      E_{RY} &= \{\sqrt{1-p_{RY}}I, \sqrt{p_{RY}}Y \}, \\
    \end{aligned}
    \end{equation}
    where in our simulation $p_{XX}=10\,p_{RX} = 10\,p_{RY}$ is the M{\o}lmer-S{\o}rensen control error rate.\\
    After each entangling M{\o}lmer-S{\o}rensen gate we model a motional mode heating error with the Kraus channel
    \begin{equation}
        E_{heating} = \{ \sqrt{1-p_h}II, \sqrt{p_h}XX\},
    \end{equation}
    where $p_h = r_{heating} \times t_{MS}$, $r_{heating}$ is the heating rate and $t_{MS}$ is the time of the corresponding entangling M{\o}lmer-S{\o}rensen gate.\\
    After all one- and two-qubit gates we model a single-qubit dephasing error on each qubit involved in the gate with the Kraus channel
    \begin{equation}
        E_{dephasing} = \{ \sqrt{1-p_d}I, \sqrt{p_d}Z\},
    \end{equation}
    where $p_d = r_d \times t_g$, $r_d$ is the dephasing rate and $t_g$ is the time of the applied gate.
\end{enumerate}

\subsection{Ion trap operation times}
\begin{figure}
\centering
    \begin{subfigure}[b]{0.24\textwidth}
        \centering
        \includegraphics[width=\linewidth]{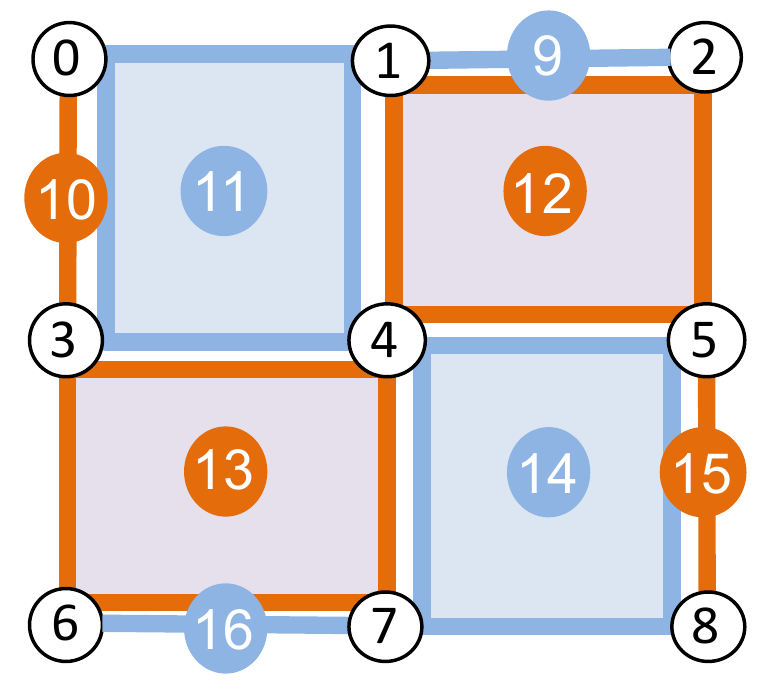}
        \caption{Surface-17}
    \end{subfigure}%
    \hfill%
    \begin{subfigure}[b]{0.233\textwidth}
        \centering
        \includegraphics[width=\linewidth]{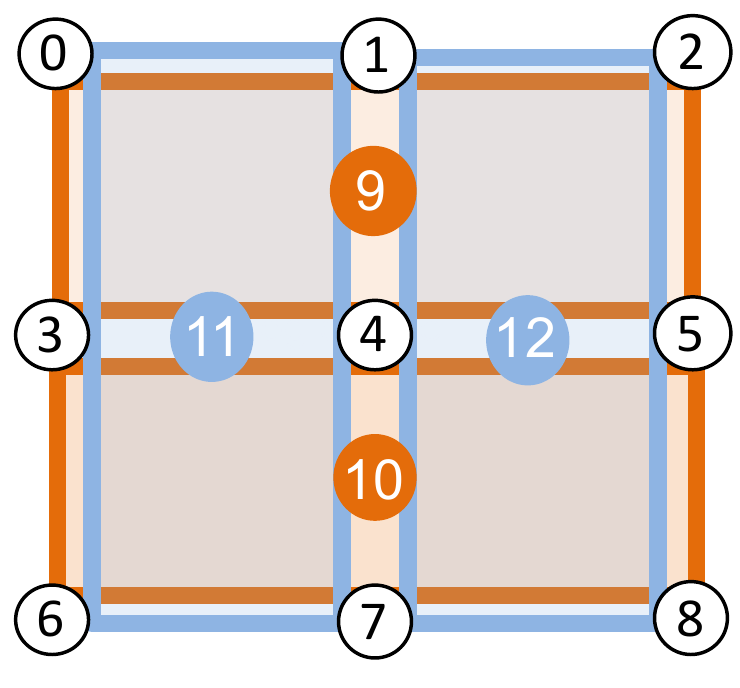}
        \caption{Bacon-Shor-13}
    \end{subfigure}
\captionsetup{justification=raggedright}
\caption{Qubit labeling for the [[9,1,3]] surface code and Bacon-Shor code. Data qubits, ancillary qubits for $X$ type stabilizers, and ancillary qubits for $Z$ type stabilizers are shown in black, blue, and orange circles, respectively.}
\label{fig:iontostabs}
\end{figure}
Here we present some results on trap operation times for ion arrangements optimized according to different parameters. More details of the ion trap model and justification for the parameters are available in Ref. \cite{trout2017simulating}.
The optimization takes into account that although 2-qubit gates can occur between any pair of ions, the gate time will depend on the ion distance.  It also assumes that before the ancillary qubits are measured they must be physically displaced from the data qubits. If the data qubits and ancillary qubits are separated in the chain, the number of joining and splitting steps is reduced at the cost of longer logical gate times.  We note that neither limit is fundamental.  It is possible to make 2-qubit gates where there is no time dependence on ion distance by adjusting laser power \cite{LeungFMgates2017} and that ancillary qubits do not need to be separated when two ion species are used \cite{BermudezITQCwithSteane2017}.

The optimization was done using a simulated annealing algorithm with an objective function adjusted for different optimization parameters, total time assuming two parallel gates (T) or average 2-qubit gate time (A), and different constraints, mixed (M) or separated (S) data and ancillary qubits. For each optimization label we calculate times for operations performed in both serial and parallel. Parallel operations allow for two simultaneous two-qubit gates exciting the independent $x$ and $y$ radial modes and fully parallel single-ion operations. We assume single-qubit gates, parallel measurement/state preparation, and shuttling between operation and measurement zones require 10~$\mu$s, 100~$\mu$s, and 100~$\mu$s, respectively. The shuttling operation includes the time to split and join ion chains. The results in the paper are for the MA optimization, which minimizes error for the error correction step.  In Ref. \cite{trout2017simulating}, the reported results for Surface-17 are for MT.\\

\begin{table}[H]
\begin{center}
\begin{tabular}{ c|c|c|c|c|c }
 \hline
 Code & Opt. & Logic & Shuttle & Meas. & Total \\ \hline \hline
\multirow{3}{*}{Surface-17} & SA & 7240 (3920) & 200 & 100 & 7950 (4220) \\
 & MA & 2300 (1170) & 1800 & 800 & 4900 (3770) \\ 
 & MT & 4300 (2320) & 700 & 300 & 5300 (3320) \\ \hline
\multirow{3}{*}{Bacon-Shor-13} & SA & 5580 (3270) & 200 & 100 & 5880 (3570) \\
 & MA & 2910 (1490) & 1000 & 400 & 4310 (2890) \\
 & MT & 3580 (1860) & 400 & 100 & 4080 (2360) \\ \hline
\end{tabular}
\captionsetup{justification=raggedright}
\caption{Trap operation times (in $\mu$s) for one round of error correction calculated according to ion arrangements optimized for an array of parameters (see text). All values are reported in $\mu$s and the numbers in parentheses refer to the gate time when two 2-qubit operations are performed in parallel.}
\label{table:opts}
\end{center} 
\end{table}

\begin{table}[H]
\begin{center}
\begin{tabular}{ c|c|c }
 \hline
 Code & Opt. & Prep Time \\ \hline \hline
\multirow{3}{*}{Surface-17} & SA & 7880-11820 (4460-6690) \\
 & MA & 2400-3600 (1200-1800) \\ 
 & MT & 3800-5700 (2100-3150)\\ \hline
\multirow{3}{*}{Bacon-Shor-13} & SA & 670 (450) \\
 & MA & 1670 (1040)\\
 & MT & 1480 (970)\\ \hline
\end{tabular}
\captionsetup{justification=raggedright}
\caption{Trap operation times (in $\mu$s) for logical state preparation calculated according to ion arrangements optimized for an array of parameters (see text). The time for Surface-17 can vary as it is a probabilistic circuit of syndrome extraction.}
\label{table:prep}
\end{center} 
\end{table}

\end{document}